\documentclass[aps,prd,floatfix,superscriptaddress,preprintnumbers,amsmath,amssymb]{revtex4}
\usepackage{amssymb}
\usepackage{amsmath}
\usepackage{amsfonts}
\usepackage{epsfig}
\usepackage{graphicx}
\usepackage{tabularx}
\usepackage{color}

\newcommand{\seq}{\begin{subequations}}
\newcommand{\sen}{\end{subequations}}
\newcommand{\eq}{\begin{eqnarray}}
\newcommand{\en}{\end{eqnarray}}

\def\shiftdown#1{#1\llap{\lower.04ex\hbox{#1}}}

\newcommand{\ra}{\rangle}
\newcommand{\la}{\langle}

\def\arraystretch{1.5}

\begin{document}

\title{Mesons in a soft-wall AdS-Schwarzschild approach at low temperature} 

\author{Thomas Gutsche}
\affiliation{Institut f\"ur Theoretische Physik,
Universit\"at T\"ubingen, \\
Kepler Center for Astro and Particle Physics,
Auf der Morgenstelle 14, D-72076 T\"ubingen, Germany}
\author{Valery E. Lyubovitskij}
\affiliation{Institut f\"ur Theoretische Physik,
Universit\"at T\"ubingen, \\
Kepler Center for Astro and Particle Physics,
Auf der Morgenstelle 14, D-72076 T\"ubingen, Germany}
\affiliation{Departamento de F\'\i sica y Centro Cient\'\i fico
Tecnol\'ogico de Valpara\'\i so-CCTVal, Universidad T\'ecnica
Federico Santa Mar\'\i a, Casilla 110-V, Valpara\'\i so, Chile}
\affiliation{Department of Physics, Tomsk State University,
634050 Tomsk, Russia}
\affiliation{Laboratory of Particle Physics, 
Tomsk Polytechnic University, 634050 Tomsk, Russia} 
\author{Ivan Schmidt}
\affiliation{Departamento de F\'\i sica y Centro Cient\'\i fico
Tecnol\'ogico de Valpara\'\i so-CCTVal, Universidad T\'ecnica
Federico Santa Mar\'\i a, Casilla 110-V, Valpara\'\i so, Chile}
\author{Andrey Yu. Trifonov}
\affiliation{Laboratory of Particle Physics, 
Tomsk Polytechnic University, 634050 Tomsk, Russia} 

\vspace*{.2cm}

\date{\today}

\begin{abstract}

We derive a holographic soft-wall approach in five dimensional 
AdS-Schwarzschild space for the description of mesons at finite temperature.  
In this first application we consider the small temperature limit and derive 
analytical expression for the mass spectrum of mesons 
with adjustable quantum numbers $n$ (radial number), 
$L$ (angular orbital momentum) and $J$ (total angular momentum). 
We explicitly separate the contribution at zero temperature and the
leading order temperature correction. The temperature corrections 
arise from the temperature dependence of the dilaton parameter 
(which is the parameter of spontaneous breaking of chiral symmetry 
related to the pseudoscalar meson decay constant) and the warping of 
the AdS metric due to temperature. We extend our results 
to any hadron with integer spin (tetraquarks, dibaryons, etc.). 
We present numerical analysis for the temperature dependence of 
meson masses and form factors.  

\end{abstract}

\maketitle

\section{Introduction}

The study of hadron properties at finite temperature is a promising task, 
since it allows for a deeper understanding of the evolution of the 
early Universe, the formation of hadronic matter and its phase transitions. 
Hadrons at finite temperature have been 
considered in the past in holographic QCD (see, e.g., 
Refs.~\cite{Herzog:2006ra}-\cite{Chen:2018jha}). In particular, 
In Ref.~\cite{Herzog:2006ra} it was shown that the deconfinement 
in the anti-de Sitter approaches 
of quantum chromodynamics (AdS/QCD) occurs via a first-order Hawking-Page 
phase transition between low temperature thermal AdS and high temperature 
black hole. The model-dependent predictions for the deconfinement 
temperature in the hard (HW) and soft-wall (SW) AdS/QCD have been obtained: 
$T_c^{\rm HW} = 122$ MeV and $T_c^{\rm SW} = 191$ MeV, where the SW 
prediction was close to the lattice QCD prediction 
$T_c^{\rm Lattice} = 192 \pm 7 \pm 4$ MeV~\cite{Cheng:2006qk}. 
A better prediction of the SW AdS/QCD for the deconfinement temperature 
was related to a more realistic description of meson spectrum in this 
approach. In Ref.~\cite{Grigoryan:2010pj} the gravity dual of charmonium 
in the strongly coupled QCD plasma was constructed on a basis of the SW 
AdS/QCD model and performing the matching of the ultraviolet 
behavior of the charm current correlator to the result of QCD. 
Detailed analysis of the holographic potential forming the charmonium 
and its spectral function has been presented. 
In Ref.~\cite{Colangelo:2010pe} the free energy of the heavy 
quark-antiquark pair in strongly interaction matter has been investigated 
using a holographic approach formulated with the use of the 
AdS/Reissner-Nordstr\"om black-hole metric at finite temperature $T$ 
and chemical potential $\mu$. The obtained deconfinement line in the 
$\mu-T$ plane was similar to ones obtained in lattice QCD and 
other QCD related approaches. In Ref.~\cite{Colangelo:2011sr}
a SW AdS/QCD approach with the AdS-Schwarzschild geometry 
has been applied for study the dependence of 
the chiral condensate on the temperature and quark density. 
The main finding was that for $\mu$ below a critical value the 
chiral quark condensate is decreasing with increasing the temperature.  
In particular, the quark condensate at $\mu = 0$ vanishes at 
$T_c \simeq 210$ MeV. Increasing of $\mu$ leads to decreasing of $T_c$. 
In Ref.~\cite{Colangelo:2012jy} spectral functions of the scalar glueball 
and light vector mesons have been studied in a hot dense medium 
in a SW AdS/QCD approach based on AdS/Reissner-Nordstr\"om metric. 
It was observed that scalar glueball and vector mesons became 
unstable at increased values of temperature and chemical potential. 
Similar analysis of the vector mesons in the SW AdS/QCD approach based 
using the AdS-Schwarzschild geometry has been performed 
in Ref.~\cite{Mamani:2013ssa}. In Refs.~\cite{Braga:2017bml}
detailed study of thermal properties of charmonium and bottomonium 
vector mesons has been carried out in holographic AdS/QCD 
The dissociation of heavy vector quarkonia states 
was studied in the context of the configurational entropy (CE) setup. 
It was found that CE has a specific behavior on temperature 
for charmonium and bottomonium states. In particular, for the 
charmonium the CE curve increases monotonically 
and the probability of dissociation of these states 
in the medium increases with the temperature, while for the bottomonium 
the picture is different. In case of the bottomonium, 
the CE has global minimum at temperatures $T \sim 1.3 T_c$ 
and then for higher temperatures it increases monotonically. 
In Ref.~\cite{Bartz:2016ufc} in-medium properties of mesons have been 
studied at finite temperature and baryon chemical potential within 
SW AdS/QCD model with modified dilaton field and holographic 
potential (by adding the quartic scalar term) 
in order to obtain the correct form of chiral symmetry breaking 
and correct spectrum. In particular, it was used 
a dilaton field parametrization, which is negative quadratic in 
the ultraviolet limit, while becoming positive quadratic in the 
infrared region. In Ref.~\cite{Vega:2017dbt} different types of the SW 
AdS/QCD potentials have been analyzed in order to obtain a melting 
temperature for different bound states --- scalar mesons, glueballs, 
hybrids, and tetraquarks. One of the main findings was an observation 
that the melting temperature increases for hadrons containing heavy 
quarks. In Ref.~\cite{Vega:2018dgk} it was proposed an idea of 
a thermal dilaton --- dilaton field depending on temperature. 
Two thermal forms of the dilatons have been studied which make it 
possible to obtain melting temperatures for mesons close to 180 MeV. 
Detailed analysis of the deconfinement temperature 
for glueballs, scalar and vector mesons in different versions of the 
HW and SW AdS/QCD approaches has been performed in Ref.~\cite{Afonin:2014jha}.
In Ref.~\cite{Afonin:2015fga} the SW AdS/QCD model has been applied for 
the description of the high-$T_c$ superconductivity. 
In Ref.~\cite{Chen:2018jha} chiral phase transition has been studied 
in a SW AdS/QCD model using AdS-Schwarzschild metric and 
incorporating $SU_L(N_f) \times SU_R(N_f)$ symmetry. 
In particular, a detailed analysis of the chiral condensate on 
temperature has been carried out. 

The main motivation for our study is to propose the modification 
of the soft-wall model at finite temperature in order bring it in 
consistency with QCD. In particular, we argue that in order to 
reproduce a temperature behavior of quark condensate one should include 
temperature dependence of the dilaton field, which is the parameter of 
spontaneous breaking of chiral symmetry related to the pseudoscalar meson 
decay constant and the warping of the AdS metric due to temperature. 
In particular, we propose that the dilaton field has the specific 
$T$-dependence, which is dictated by the temperature behavior 
of the chiral quark condensate in QCD~\cite{Gasser:1986vb,Leutwyler:1987th} 
derived using chiral perturbation theory (ChPT)~\cite{ChPT}.
In this way we postulate the temperature dependence of the 
dilaton field using its relation to the chiral quark quark at 
zero temperature. We note that a thermal behavior of the dilaton 
has been proposed before in Ref.~\cite{Vega:2018dgk}, but now 
in our paper we do it in a way consistent with QCD. It makes such study 
important to improve understanding hadron properties at finite temperatures. 

In the present paper we are interested in the specific low temperature 
limit and the derivation of analytical formulas for the mass spectrum of 
mesons and their form factors. In particular, we consider two possible 
sources of temperature dependence: 
(1) the warping of the AdS metric due to temperature, 
(2) the temperature dependence of the dilaton-background field, 
which produces confinement and is responsible for the breaking of 
conformal invariance and the spontaneous breaking of chiral symmetry 
in holographic QCD. 

We consider the propagation of a meson field $M_J(x,z)$ with total 
angular momentum $J$ in five dimensional AdS-Schwarzschild space at 
finite temperature. The AdS-Schwarzschild metric is specified by
\eq
ds^2 = e^{2 A(z)} \, 
\biggl[ f_T(z) dt^2 - (d\vec{x})^2 - \frac{dz^2}{f_T(z)} \biggr]
\en
where $x=(t,\vec{x}\,)$ is the set of Minkowski coordinates, 
$z$ is the holograpchic coordinate, $R$ is the AdS radius and 
$A(z) = \log(R/z)$. 
Here $f_T(z) = 1 - z^4/z_H^4$ 
where $z_H$ is the position of the event horizon, which is related to 
the black-hole Hawking temperature $T = 1/(\pi z_H)$. 
The latter also
represents (holography correspondence) the temperature of the boundary
field theory. The holographic coordinate changes from 0 to $z_H$. 
The AdS-Schwarzschild metric breaks conformal invariance because of the
temperature dependence. At zero temperature the conformal invariance can be 
broken by introducing a wall in the $z$-direction. In the following we 
consider one of the versions of the soft-wall AdS/QCD 
model~\cite{Karch:2006pv,Brodsky:2006uqa}. It was developed in 
Refs.~\cite{Branz:2010ub}-\cite{Gutsche:2017oro} for the study of mesons, 
baryons and exotic states with adjustable quantum numbers of total 
angular spin $J$, angular orbital momentum $L$ and radial quantum 
number $n$. Following Ref.~\cite{Karch:2006pv} we introduce 
the exponential prefactor $\exp[-\varphi(z)]$ in the effective action, 
containing the background (dilaton) field $\varphi(z) = \kappa^2 z^2$, 
where $\kappa$ is a scale parameter of the order of a few hundred MeV. 
This dilaton field breaks conformal invariance, produces confinement 
and is responsible for the  spontaneous breaking of chiral symmetry in
holographic QCD. In addition to dilaton we introduce in the action the 
thermal prefactor 
\eq\label{lambda_Tz} 
e^{-\lambda_T(z)}, \quad \lambda_T(z) = \alpha \frac{z^2}{z_H^2} 
+ \gamma \frac{z^4}{z_H^4} + \xi \frac{\kappa^2 z^6}{z_H^4} \,, 
\en 
where dimensionless parameters $\alpha$, $\gamma$, and $\xi$ 
parametrize the $z^2$, $z^4$, and $z^6$ thermal corrections. 
Later we will show that the 
parameter $\gamma$ is fixed to guarantee the gauge 
invariance and massless ground-state pseudoscalar mesons $(\pi$, 
$K$, $\eta$) in chiral limit, while the parameter $\xi$ is 
fixed to drop the radial dependence of six power in the 
holographic potential in our approach. 
Also we will demonstrate that the parameter $\alpha$ 
encodes the contribution of gravity
to the restoration of chiral symmetry at a critical temperature $T_c$.
From our analysis of this phenomena we will find (as will be seen 
below) that the $\alpha$ parameter increases the $T_c$ and 
its value should be relatively small. A limitation to small $\alpha$ 
is also consistent with small temperature limit. 
In the limit $z \ll z_H$ the thermal factors $f_T(z)$ and 
$\lambda_T(z)$ reduce to 1,   
and the AdS-Schwarzschild geometry reduces to the pure AdS case:
\eq
ds^2 = e^{2 A(z)} \
\biggl[dt^2 - (d\vec{x})^2 - dz^2 \biggr] \,. 
\en
Therefore, there are two sources for the breaking of 
conformal invariance in the soft-wall AdS/QCD model: 
the dilaton field $\varphi(z)$ and the metric (warping factor 
$f_T(z)$ and prefactor $\exp[-\lambda_T(z)]$). 

Recently, in Ref.~\cite{Vega:2018dgk} it was proposed to include an
additional temperature dependence of the soft-wall AdS/QCD action 
via an explicit $T$-dependence of the dilaton field: 
$\varphi(z) \to \varphi(z,T) = \kappa^2(T) z^2$. In particular,  
two forms for the thermal dilaton were proposed 
for $\varphi(z,T)$ behavior~\cite{Vega:2018dgk}: 
\eq 
\varphi_1(z,T) = \kappa^2 (1 + \alpha T)  z^2 
\en
and 
\eq 
\varphi_2(z,T) = \kappa^2_1 (1 + \alpha T) z^2 \  
{\rm tanh}(\kappa^2_2 (1 + \alpha T) z^2) \,.
\en 

Here we also propose a specific $T$-dependence of the dilaton scaling 
parameter $\kappa$. We base this choice on the idea that $\kappa^2$
as a parameter of spontaneous breaking of chiral symmetry is related 
in the soft-wall AdS/QCD approach to the quark condensate. 
At zero temperature one has the definition 
\eq\label{Sigma0}
\Sigma = \la 0 |\bar q q| 0 \ra = - N_f B \, F^2 \,,
\en 
where $N_f$ is the number of quark flavors, $B$ is the 
quark condensate parameter, and $F$ is the pseudoscalar 
meson decay constant in the chiral limit at zero temperature (e.g., 
$F \simeq 87$ MeV for $N_f=2$~\cite{Gasser:1986vb}). 
In particular, in the soft-wall AdS/QCD model the dilaton parameter 
$\kappa$ and the decay constant $F$ are related in the chiral limit 
at zero temperature as~\cite{Brodsky:2006uqa,Branz:2010ub}: 
\eq\label{F0} 
F = \kappa \frac{\sqrt{3}}{8} \,. 
\en  
Substituting Eq.~(\ref{F0}) into Eq.~(\ref{Sigma0}) we 
get a relation between $\Sigma$ and $\kappa^2$: 
\eq\label{Sigma_kappa} 
\Sigma = - \frac{3 N_f B}{64} \, \kappa^2 \,. 
\en 
We suppose identical temperature dependence of the 
$\kappa^2(T)$ and $\Sigma(T)=\la 0|\bar q q|0\ra_T$ 
postulating the relation  
\eq\label{kappaT} 
\kappa^2(T) = \kappa^2 \, \frac{\Sigma(T)}{\Sigma} \,. 
\en  
Taking into account Eqs.~(\ref{Sigma0}), (\ref{F0}), 
and the relation 
\eq\label{defBT} 
\Sigma(T) = - N_f \, B(T) \, F^2(T)  
\en 
we can relate $\kappa^2(T)$ with $T$-dependent 
quark condensate parameter $B(T)$ 
and pseudoscalar coupling constant in the chiral limit $F(T)$ as 
\eq 
\kappa^2(T) = \kappa^2 \, \frac{B(T)}{B} \, \frac{F^2(T)}{F^2} 
            = \frac{64}{3} \, \frac{B(T)}{B} \, F^2(T) \,. 
\en  
In Ref.~\cite{Gasser:1986vb}, using two-loop chiral perturbation 
theory (ChPT), the low-temperature dependence of the 
quark condensate $\Sigma(T)$ was established: 
\eq\label{SigmaT} 
\Sigma(T) &=& \Sigma \, \biggl[ 
1 - \frac{N_f^2-1}{N_f} \, \frac{T^2}{12 F^2} 
- \frac{N_f^2-1}{2 N_f^2} \, \biggl(\frac{T^2}{12 F^2}\biggr)^2 
+ {\cal O}\big(T^6\big) \biggr]= 
\Sigma \, \biggl[ 1 + \Delta_T + {\cal O}\big(T^6\big) \biggr] \,.
\en  
This result is valid for an adjustable number of quark flavors with 
$N_f \ge 2$ and is given as an expansion in $T^2$. 

The temperature correction to the condensate, up to order $T^4$, is encoded 
in the quantity $\Delta_T$ with 
\eq 
\Delta_T &=& \delta_{T_1} \frac{T^2}{12 F^2}
\,+\, \delta_{T_2} \biggl(\frac{T^2}{12 F^2}\biggr)^2 \,, 
\nonumber\\
\delta_{T_1}  &=&  - \frac{N_f^2-1}{N_f}     \,, \quad  
\delta_{T_2} \,=\, - \frac{N_f^2-1}{2 N_f^2} \,. 
\en  
In Ref.~\cite{Leutwyler:1987th} the two-loop ChPT result of 
Ref.~\cite{Gasser:1986vb} has been extended to the three-loop case  
by the inclusion of the higher-order ${\cal O}(T^6)$ term: 
\eq
\Sigma(T) &=& 
\Sigma \, \biggl[ 1 + \tilde\Delta_T + {\cal O}\big(T^8\big) \biggr] \,,
\nonumber\\
\tilde\Delta_T &=& \Delta_T + N_f (N_f^2-1) 
\biggl(\frac{T^2}{12 F^2}\biggr)^3 \, {\rm log}\frac{T}{\Lambda_\Sigma}\,. 
\en 
The scale $\Lambda_\Sigma$ absorbs the ultraviolet divergencies 
in the three-loop graphs, generated by the leading term 
${\cal} L^{(2)}_{\rm ChPT}$ in the ChPT Lagrangian. The value of 
$\Lambda_\Sigma$ can been fixed by the low-energy constants in the 
next-to-leading ChPT Lagrangian ${\cal L}^{(4)}_{\rm ChPT}$ using data. 
E.g., in the two-flavor case $\Lambda_\Sigma$ has been related to 
the $D$-wave isospin zero $\pi\pi$ scattering length $a_2^0$, leading to 
$\Lambda_\Sigma = 470 \pm 110$ MeV. Here we restrict to an accuracy of 
${\cal O}(T^4)$ in the temperature expansion of the quark 
condensate. Therefore, using the relations~(\ref{kappaT}) 
and~(\ref{SigmaT}) we obtain the $T$-dependence of the dilaton field with 
\eq
\varphi(z,T) &=& \kappa^2(T) z^2\,, \quad\quad  
\kappa^2(T) = \kappa^2 \biggl[1 \,+\, \Delta_T  
\,+\,{\cal O}\big(T^6\big)\biggr] \,. 
\en 
In the following it is useful to combine the two terms, the $T$-dependent 
dilaton field and the $z^2$ term in the thermal prefactor 
$e^{-\lambda_T(z)}$ (\ref{lambda_Tz}) as 
\eq\label{varphiZT} 
\varphi(z,T) + \frac{\alpha z^2}{z_H^2} = K^2_T z^2\,,
\en 
where 
\eq\label{def_rho}
K^2_T = \kappa^2(T) + \frac{\alpha}{z_H^2} = (1 + \rho_T) \, \kappa^2\,, 
\quad 
\rho_T = \biggl(\frac{9 \alpha \pi^2}{16}
\,+\, \delta_{T_1}\biggr) \frac{T^2}{12 F^2}
\,+\, \delta_{T_2} \biggl(\frac{T^2}{12 F^2}\biggr)^2
\,+\,{\cal O}(T^6) \,.
\en 
Note the $T$-dependence of $B(T)$ and $F(T)$ has been studied 
in Refs.~\cite{Gasser:1986vb} and~\cite{Toublan:1997rr}.    
In particular, 
$F(T)$ was calculated at one-loop in Ref.~\cite{Gasser:1986vb}   
\eq\label{FT1loop}
F(T) = F \biggl[1 - \frac{N_f}{2}  \frac{T^2}{12 F^2} \,+\, 
{\cal O}(T^4) \biggr]
\en 
and at the level of two loops for $N_f = 2$ in Ref.~\cite{Toublan:1997rr} 
\eq\label{FT2loop}
F^2(T) = F^2 \biggl[1 - \frac{T^2}{6 F^2} 
+ \frac{T^4}{36 F^4} {\rm log}\frac{\Lambda_F}{T}
\,+\, {\cal O}(T^6) \biggr]\,,
\en 
where $\Lambda_F = 2.3$ GeV is the scale 
absorbing the ultraviolet divergencies. 

In the finite temperature case it is useful to introduce 
the Regge-Wheeler (RW) tortoise coordinate $r^*$ instead 
of the holographic variable $z$ via the 
substitution~\cite{Regge:1957td,Horowitz:1999jd}:  
\eq 
r^* = - \int \frac{dz}{f_T(z)} = \frac{z_H}{2} \, 
\biggl[ - {\rm arctan}\frac{z}{z_H} 
+ \frac{1}{2} \, \log\frac{1-z/z_H}{1+z/z_H}\biggr] \,.
\en 
For convenience we will use the variable $r = - r^*$. 
Note that in the low temperature limit the $r$ coordinate 
can be expanded as 
\eq 
r = z \biggl[ 1 + \frac{t^4}{5} + \frac{t^8}{9} 
+ {\cal O}\Big(t^{12}\Big)  \biggr]\,, \quad t = z/z_H \,. 
\en  
The holographic coordinate, expanded in powers of $r$, is 
\eq\label{expansion_r} 
z = r \biggl[ 1 - \frac{t_r^4}{5} + \frac{4}{45} t_r^8 
+ {\cal O}\Big(t_r^{12}\Big)  \biggr]\,, \quad t_r = r/z_H \,. 
\en 
In order of accuracy, ${\cal O}(T^4)$, we are working we restrict 
to the leading-order (LO) and next-to-leading-order (NLO) term 
in the expansion of $z$ in Eq.~(\ref{expansion_r}). 
In this case the metric is 
\eq\label{metric_fT_mod}
ds^2 = e^{2 A(r)} \,  f_T^{3/5}(r) \, 
\biggl[dt^2 - \frac{d\vec{x\,}^2}{f_T(r)} - dr^2\biggr]\,, 
\quad 
A(r) = \log(R/r)\,, \quad f_T(r) = 1 - r^4/z_H^4 \,. 
\en 
The product of the prefactors containing dilaton field and 
thermal factor $\lambda_T$ in terms of the $r$ variable is 
\eq\label{calP} 
{\cal P} = \exp\Big[-\varphi_T(r) - \gamma \frac{r^4}{z_H^4} 
- \frac{\kappa^2 r^6}{z_H^4} \Big( \xi - \frac{2}{5} \Big) \Big] \,, 
\en 
where $\varphi_T(r)$ is the dilaton field 
\eq\label{dilaton_fT}  
\varphi_T(r) = K_T^2 r^2 = (1 + \rho_T) \kappa^2 r^2 \,.
\en 
In order to suppress the $r^6$ terms, which could later on enter to 
the holographic potential defining the properties of hadrons, 
we fix the parameter $\xi$ as $\xi = 2/5$. The parameter $\gamma$ 
is fixed to guarantee the gauge 
invariance and massless ground-state pseudoscalar mesons $(\pi$, 
$K$, $\eta$) in chiral limit, i.e. it should provide that 
$r^4$ terms are vanish for $J=0$ and $J=1$ and contribute only 
for higher spins states $J \ge 2$. It will be fixed 
fixed later (in next section) in the consideration of 
the equation of motion for the AdS bulk profile. 
After fixing the $\xi$ parameter, the prefactor ${\cal P}$ reads 
\eq\label{calP2}
{\cal P} = \exp\Big[-\varphi_T(r) - \gamma \frac{r^4}{z_H^4} \Big] \,.
\en
In Eqs.~(\ref{calP}) and~(\ref{calP2}) we drop higher-order temperature 
dependent terms ${\cal O}(T^6)$.

\begin{figure}
\begin{center}
\epsfig{figure=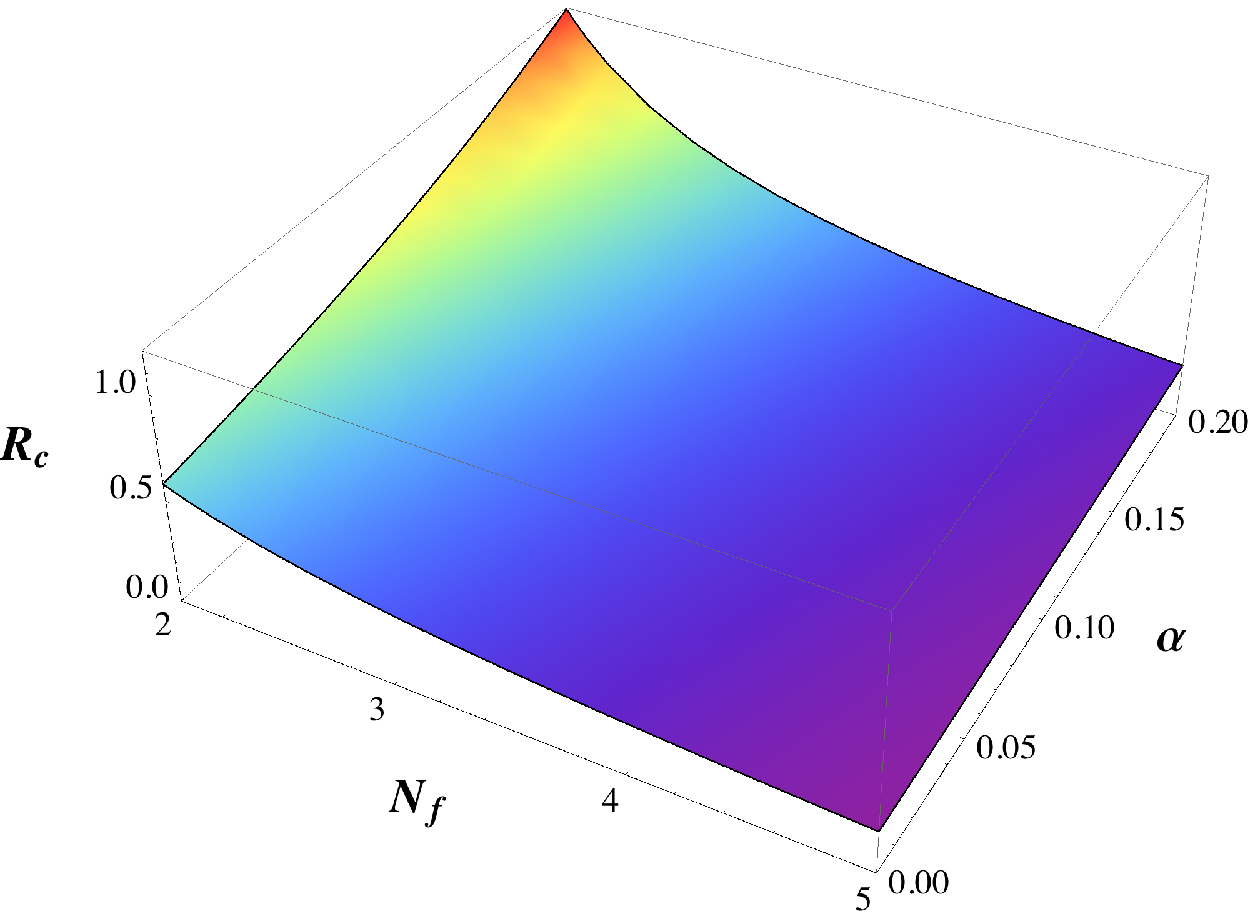,scale=.6}
\end{center}
\noindent
\caption{Dependence of the critical temperature 
(contained in $R_c = T_c^2/12F^2$) on the number 
of flavors $N_f$ and the parameter~$\alpha$. 
\label{fig:Rck}}

\centering 
\vspace*{.5cm}
\begin{minipage}{0.45\textwidth}
\centering 
\epsfig{figure=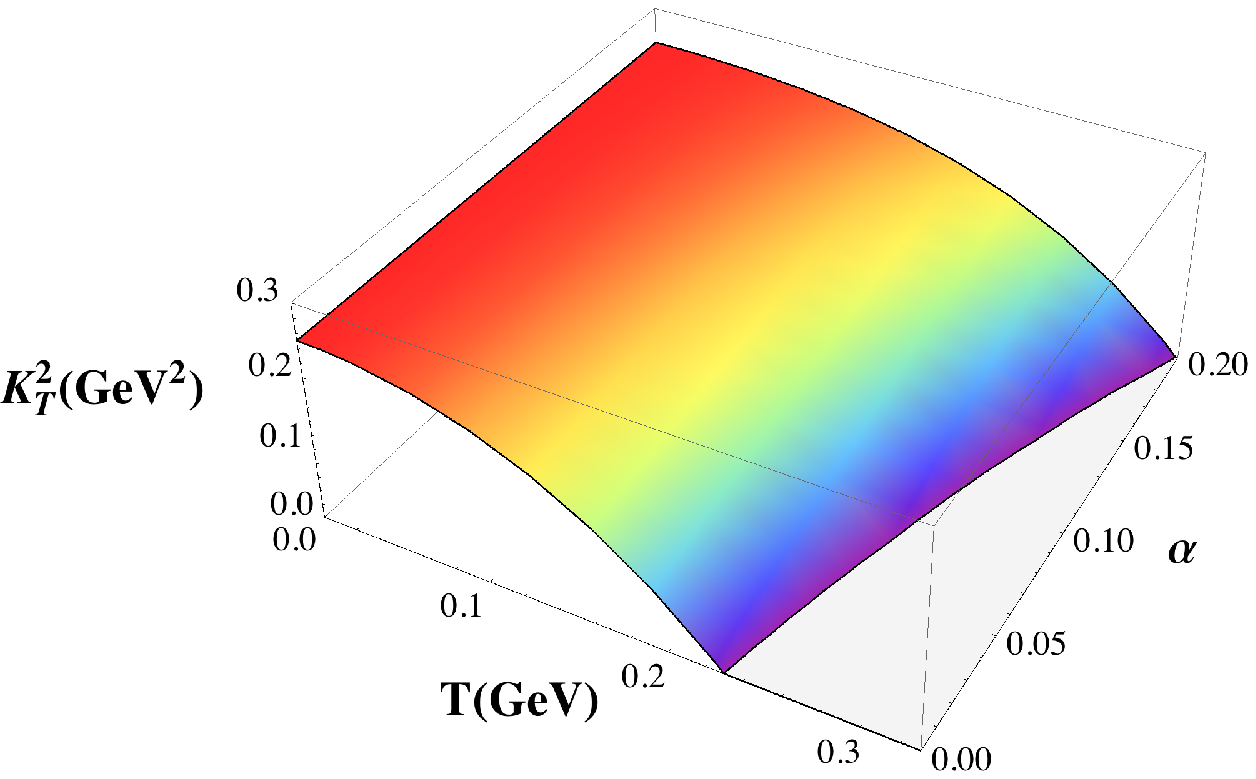,scale=0.6}
\caption{$K^2_T$ for $N_f = 2$ and $F=87$ MeV.}
{\label{fig:kappa2}} 
\end{minipage}
\begin{minipage}{0.45\textwidth}
\centering 
\epsfig{figure=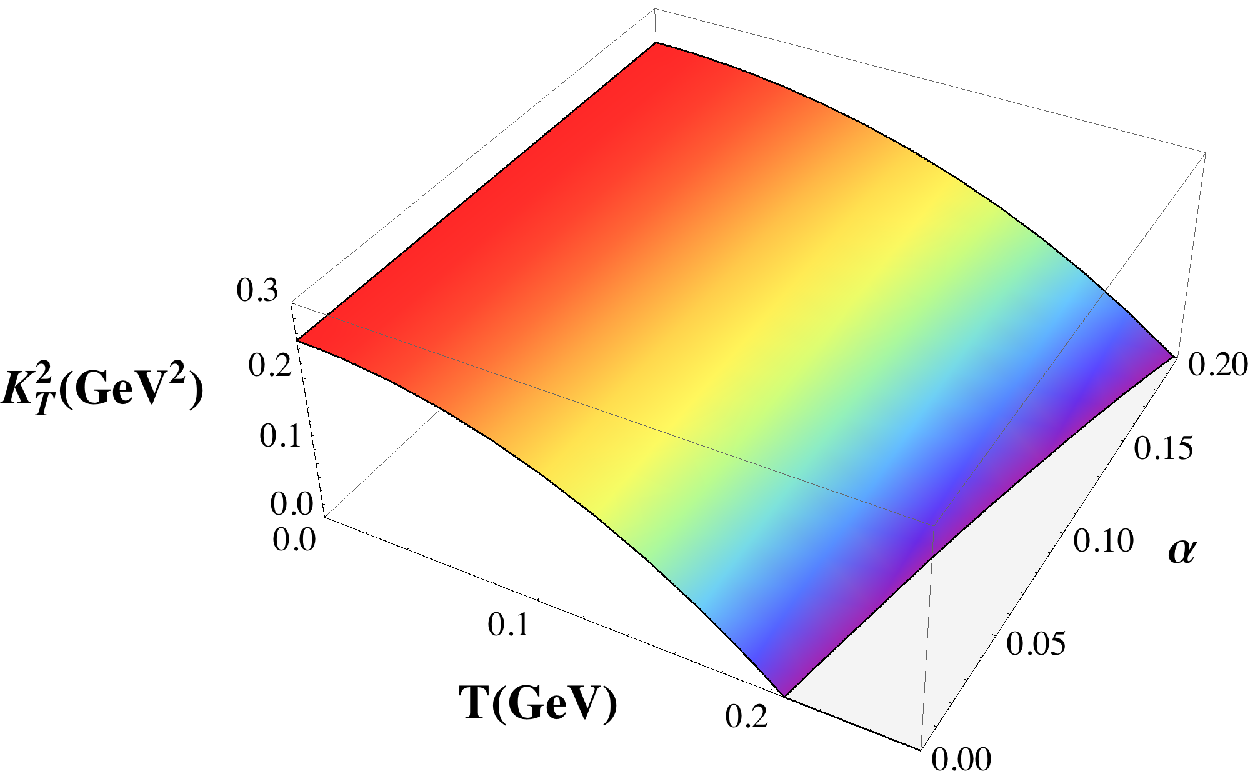,scale=0.6}
\caption{$K^2_T$ for $N_f = 3$ and $F=100$ MeV.}
{\label{fig:kappa3}} 
\end{minipage}

\centering 
\vspace*{.5cm}
\begin{minipage}{0.45\textwidth}
\centering 
\epsfig{figure=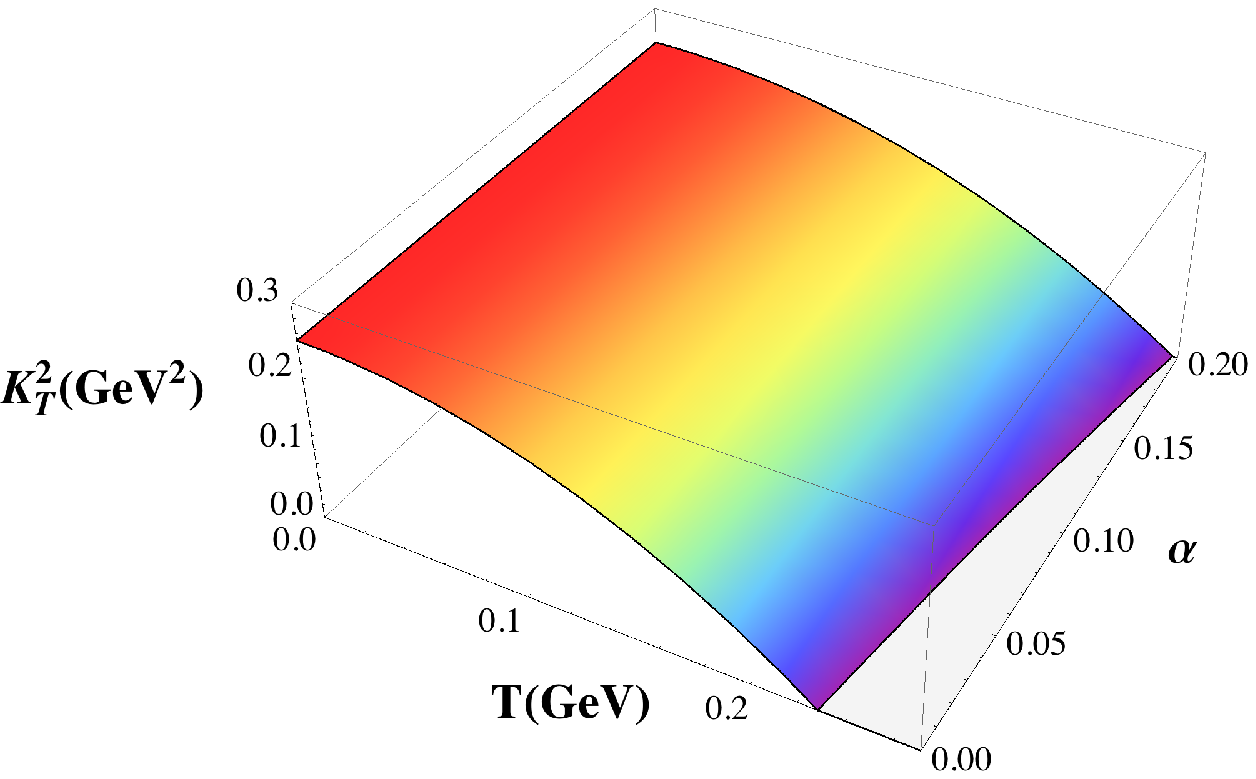,scale=0.6}
\caption{$K^2_T$ for $N_f = 4$ and $F=130$ MeV.}
{\label{fig:kappa4}} 
\end{minipage}
\begin{minipage}{0.45\textwidth}
\centering 
\epsfig{figure=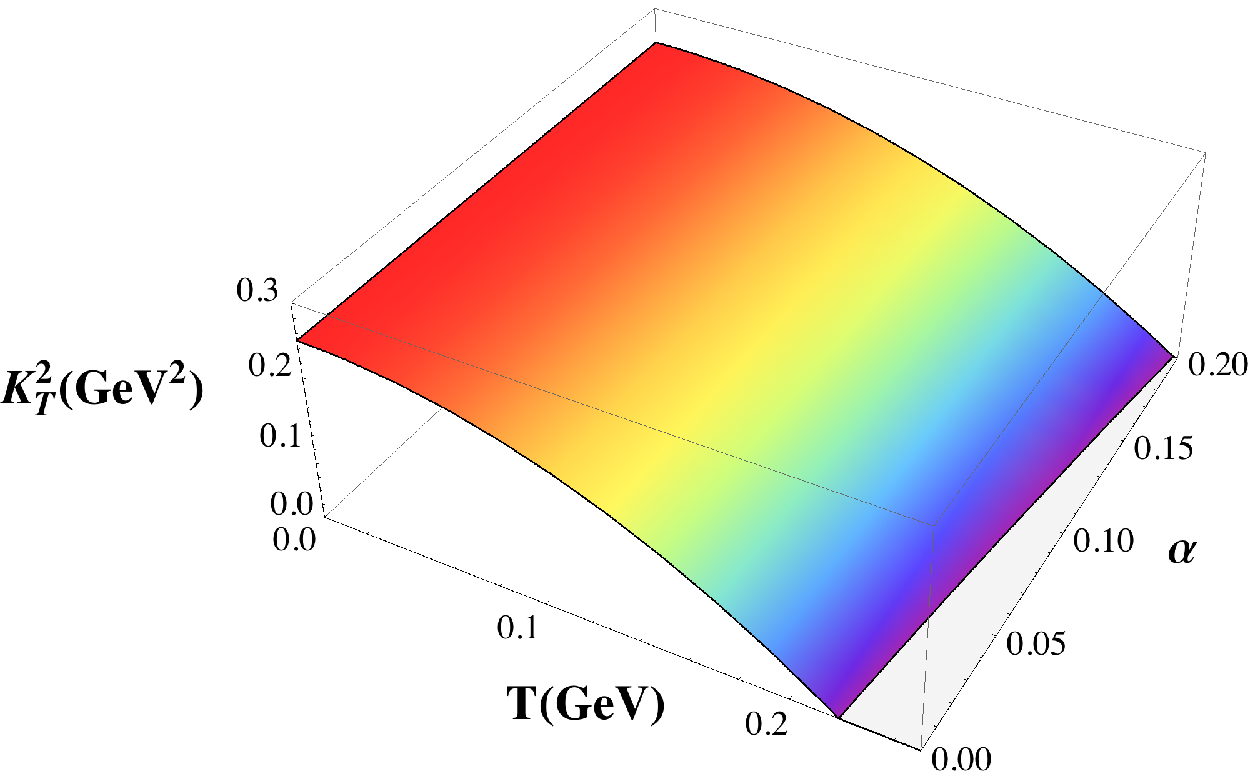,scale=0.6}
\caption{$K^2_T$ for $N_f = 5$ and $F=140$ MeV.}
{\label{fig:kappa5}} 
\end{minipage}

\end{figure}

From Eq.~(\ref{def_rho}) we can fix the 
critical temperature $T_c$ in the soft-wall AdS/QCD model  
which corresponds to a vanishing dilaton parameter $K^2_{T_c} = 0$. 
Note that in comparison with ChPT we have an additional piece contributing 
to chiral restoration, which comes from the warping thermal factor 
induced by gravity. We expect that this additional gravitational effect 
must be perturbative and does not significantly change the QCD prediction. 
In particular, QCD based on the two-loop ChPT calculation~(\ref{SigmaT})   
predicts the following result for the critical temperature at order $T^4$  
\eq 
\frac{\Big(T_c^{\rm QCD}\Big)^2}{12 F^2} = N_f \, \biggl[ 
\sqrt{\frac{N_f^2+1}{N_f^2-1}} - 1 \biggr] \,.
\en 
$T_c^{\rm QCD}$ has the power scaling behavior 
$T_c^{\rm QCD} \sim F/\sqrt{N_f}$ at large $N_f$. 
The thermal warping factor changes the QCD prediction for $T_c$ to 
\eq 
\frac{T_c^2}{12 F^2} 
= N_f \, \biggl[ 
\sqrt{\frac{N_f^2+1}{N_f^2-1} - 2 \beta + \beta^2} - 1 + \beta \biggr] \,,
\en 
where 
\eq 
\beta = \frac{9 \alpha \pi^2}{16} \, \frac{N_f}{N_f^2-1} \,. 
\en 
At large $N_f$ and small $\alpha$, $T_c$ can be expanded as 
\eq
T_c = 2F \sqrt{\frac{3}{N_f}} \, \biggl[ 1 + \frac{9 \pi^2}{32} \, 
\frac{\alpha}{N_f} \,+\,  {\cal O}(1/N_f^2)\biggr]\,. 
\en  
In Fig.~\ref{fig:Rck} we plot the dependence of the quantity 
$R_c = T_c^2/(12 F^2)$ on the number of quark flavors $N_f$ and 
the coupling~$\alpha$. One can see that $R_c$ decreases when $N_f$ 
increases, while it increases with a growth of the parameter $\alpha$. 
$R_c$ rapidly grows with increasing $\alpha$ for $N_f = 2$. 
In Figs.~\ref{fig:kappa2}-\ref{fig:kappa5} we present the 3 dimensional 
plots for the $T$-dependence of the dilaton parameter $K^2_T$ 
for four cases: 
(1) $N_f=2$, $F=87$~MeV, 
(2) $N_f=3$, $F=100$~MeV, 
(3) $N_f=4$, $F=130$~MeV, and 
(4) $N_f=5$, $F=140$~MeV. 
In each case the $T_c$ regime corresponds to the intersection of 
the 3 dimensional plot with the ($T,\alpha$) plane. 
The critical temperature for all number of flavors is enhanced 
when increasing the parameter $\alpha$. 
In Table~\ref{tab:Tc_alphaNf} we present numerical results 
for the critical temperature for different number of flavors and 
for three specific values of $\alpha = 0, 0.1, 0.2$. 
One can see that the value of the $\alpha$ parameter.  
In particular, it increases the $T_c$ and its value 
should be relatively small in order exclude its big 
contribution to the shift of the value of the $T_c$. 
Note, due to expansion of the chiral condensate in powers
of $T^2/(12 F^2)$ the analysis at small temperature can be valid at
$T < 2F \sqrt{3} \simeq 300-500$ MeV in case of number of
flavors varied from 2 to 5.

\begin{table}
   \vskip 1mm
   \centering
\caption{Dependence of $T_c$ on $N_f$ and $\alpha$.}
    \label{tab:Tc_alphaNf} 
     \def\arraystretch{1.25}
\begin{tabular}{|c|c|c|c|}
\hline
\quad $N_f$ \quad &  \multicolumn{3}{c|}{$T_c$ (MeV)} \\
\cline{2-4}
& \qquad  $\alpha = 0$   \qquad\qquad 
& \qquad  $\alpha = 0.1$   \qquad\qquad 
& \qquad  $\alpha = 0.2$ \qquad\qquad 
\\
\hline
2 & 230   & 270   & 329.3 \\
\hline
3 & 206.1 & 228.2 & 258   \\
\hline
4 & 228.9 & 246.6 & 268.8 \\
\hline
5 & 219.1 & 232.4 & 248.3 \\
\hline
\end{tabular}
\end{table}

After these preliminaries we can formulate our approach, starting 
from an effective action at finite temperature and then considering 
the applications to the mass spectrum and the form factors of mesons. 
The paper is structured as follows.
In Sec.~II we present the details for the  construction of an effective 
action at small temperatures and apply it to the calculation of 
the mass spectrum and form factors of hadrons with integer spin $J$ 
(mesons, tetraquarks, dibaryons, etc.). In Sec.~III we present numerical 
results for the mass spectrum and the form factors of mesons. 
Finally, in Sec.~IV, we summarize the results of the paper. 

\section{Framework} 

\subsection{Effective action and hadron masses at low temperatures}

In this section we start with the derivation of a 
five dimensional action for the boson bulk field 
$M_{N_1\ldots N_J}(x,r,T)$, with arbitrary total integer spin $J$ 
at small temperature $T$. Our formalism is based on the 
analogous action at zero temperature~\cite{Branz:2010ub} 
and includes the issues discussed in the previous section. 
The action reads: 
\eq\label{action_SM}
S_M &=&\frac{(-)^J}{2} \int d^4x dr \, \sqrt{g} \, 
e^{-\varphi_T(r) - \gamma r^4/z_H^4} \, 
\biggl[ \partial_N M_{N_1\ldots N_J}(x,r,T) 
\partial^N M^{N_1\ldots N_J}(x,r,T) \nonumber\\ 
&-& \Big(\mu_J^2(r,T) + V_J(r,T)\Big) \, 
M_{N_1\ldots N_J}(x,r,T) M^{N_1\ldots N_J}(x,r,T) 
\biggr] \, 
\en
where $(x,r)$ is the set of four Minkowski and holographic 
coordinates. The dilaton $\varphi_T(r)$ and metric $ds^2$ 
are specified in Eqs.~(\ref{dilaton_fT}) and~(\ref{metric_fT_mod}), 
$\sqrt{g} = (R/r)^5$, and 
\eq
V_J(r,T) = \frac{e^{-2 A(r)}}{f_T^{3/5}(r)} \, \Big[ 
\varphi^{\prime\prime}_T(r)
+ (3-2J) \, \varphi^{\prime}_T(r) A^{\prime}(r) \Big]
\en
is the dilaton potential with 
$F^{\prime}(r) = dF(r)/dr$,
$F^{\prime\prime}(r) = d^2F(r)/dr^2$ and $F=\varphi, \, A$. 

The quantity $\mu_J^2(r,T)$ is the bulk boson mass at finite 
temperature, which is related to the bulk boson mass at zero 
temperature $\mu_J^2$ as 
\eq 
\mu_J^2(r,T) = \frac{\mu_J^2}{f_T^{3/5}(r)} \,. 
\en 
As is  known, the $\mu_J^2$ is expressed in terms of the dimension 
$(\Delta)$ of the interpolating operator dual to the spin-$J$ bulk 
boson field as 
\eq
\mu_J^2 R^2 = (\Delta - J) (\Delta + J -  4)\,. 
\en 
We therefore have  
\eq
\mu_J^2(r,T) R^2 = 
\frac{1}{f_T^{3/5}(r)} \, (\Delta - J) (\Delta + J -  4)\,. 
\en 
For the case of bulk fields dual to the $N$-partonic state we get 
$\Delta = N + L$, where $L = {\rm max} \, | L_z |$ is the maximal
value of the $z$-component of the quark orbital angular momentum
in the light-front wave function~\cite{Brodsky:2006uqa}. 
For mesons, tetraquarks, and sixquarks/dibaryons we have 
$N = 2, 4,$ and $6$, respectively.  

Using the axial gauge $M_z(x,r,T) = 0$ we perform a Kaluza-Klein expansion
for the four-dimensional transverse components of the AdS fields 
\eq\label{KK_coord}
M_{\mu_1\ldots \mu_J}(x,r,T) =
\sum\limits_n \ M_{\mu_1\ldots \mu_J, n}(x) \ \Phi_{nJ}(r,T),
\en
where $n$ is the radial quantum number and $M_{\mu_1\ldots \mu_J, n}(x)$
is the tower of the Kaluza-Klein (KK) modes dual to mesons with spin $J$.  
$\Phi_{nJ}(r,T)$ are their extra-dimensional profiles (wave functions) 
depending on the temperature. 

After straightforward calculations~\cite{Gutsche:2011vb} 
one can derive the Schr\"odinger-type equation of motion 
for the profile $\phi_{nJ}(r,T) = e^{- B_T(r)/2} \Phi_{nJ}(r,T)$ 
with 
\eq\label{BTr} 
B_T(r) = \varphi_T(r) + A(r) (2J - 3) 
+ \frac{r^4}{5 z^4_H} (2J - 3 + 5 \gamma)  \,.
\en  
Here we can fix the parameter $\gamma$. In order to suppress 
the contribution of the $r^4$ term in the holographic potential 
to the mass spectrum of pseudoscalar ground state mesons 
($\pi$, $K$, $\eta$) and to the bulk-to-boundary propagator 
of the $J=1$ vector fields to guarantee the charge conservation 
we can fix $5 \gamma = J (J - 3) +3$. 
As result the $B_T(r)$ reads 
\eq
\label{BTr_fixed}
B_T(r) = \varphi_T(r) + A(r) (2J - 3)
+ \frac{r^4}{5 z^4_H} J (J - 1) 
\en 
and at $J=0$ and $J=1$ 
the $r^4$ term vanishes.
In the rest frame of the AdS field with $\vec{p} = 0$ we get  
\eq\label{Eq1}
\Big[ - \frac{d^2}{dr^2} \, + \, U_J(r,T) \, 
\Big] \phi_{nJ}(r,T) = M^2_{nJ}(T) \phi_{nJ}(r,T) \,, 
\en
where $U_J(r,T)$ is the effective potential at finite temperature, 
which can be decomposed into a zero temperature term 
$U_J(r) \equiv U_J(r,0)$ and a temperature dependent term
$\Delta U_J(r,T)$ 
\eq 
U_J(r,T) &=& U_J(r) \,+\, \Delta U_J(r,T)
\,, \nonumber\\
U_J(r) &=& \kappa^4 r^2 \, + \, 2 \kappa^2 (J-1) \, + \, 
\frac{4 m^2 - 1}{4 r^2} \,, 
\nonumber\\
\Delta U_J(r,T) &=& 2 \rho_T \kappa^2 \, (
\kappa^2 r^2 + J - 1 ) 
+ \frac{4 r^2}{5 z_H^4} \, J (J - 1) ( \kappa^2 r^2 - J ) \,,
\en 
where $m = N + L - 2$.  

At zero temperature $T= 0$ the Schr\"odinger-type EOM 
\eq\label{Eq2}
\Big[ - \frac{d^2}{dr^2} + U_J(r,0)
\Big] \phi_{nJ}(r,0) = M^2_{nJ}(r,0) \phi_{nJ}(r,0)
\en
has analytical solutions. The resulting wave function 
\eq\label{phi_r0}
\phi_{nJ}(r,0) = \sqrt{\frac{2 \Gamma(n+1)}{\Gamma(n+m+1)}} 
\ \kappa^{m+1}
\ r^{m+1/2} \ e^{-\kappa^2 r^2/2} \ L_n^m(\kappa^2r^2)
\en 
corresponds to the mass spectrum 
\eq\label{mass2_T0}
M^2_{nJ}(0) = 4 \kappa^2 \Big( n + \frac{m + J}{2} \Big) 
\en
of the bosonic hadrons composed of $N$ constituents with spin $J$, 
angular orbital momentum $L$, and the radial quantum number $n$. 
Here we use the generalized Laguerre polynomials
\eq
L_n^m(x) = \frac{x^{-m} e^x}{n!}
\, \frac{d^n}{dx^n} \Big( e^{-x} x^{m+n} \Big) \,.
\en
Temperature corrections to hadronic mass spectrum 
are evaluated perturbatively considering hadronic wave functions 
at $T=0$ as unperturbed solutions: 
\eq 
\Delta  M^2_{nJ}(T) = \langle \phi_{nJ}(0) | 
\Delta U_J(T) | \phi_{nJ}(0) \rangle 
= \int\limits_0^\infty dr \phi_{nJ}^2(r,0) \Delta U_J(r,T)
\en 
In the low temperature case the hadronic mass spectrum is 
\eq\label{mass2_Tfinite}
M^2_{nJ}(T) &=& M^2_{nJ}(0) \,+\, \Delta  M^2_{nJ}(T) \,, 
\nonumber\\
\Delta  M^2_{nJ}(T) &=&  \rho_T \, M^2_{nJ}(0) 
+ R_{nJ} \frac{\pi^4 T^4}{\kappa^2} \,, \nonumber\\
R_{nJ} &=& \frac{4}{5} J (J-1)\Big[ 
(m + 1) (m + 2) + (6 n - J) (n + m + 1) - n J \Big] \,. 
\en 
The solution for the bulk profile $\phi_{nJ}(r,T)$ reads 
\eq\label{phi_rT}
\phi_{nJ}(r,T) = \sqrt{\frac{2 \Gamma(n+1)}{\Gamma(n+m+1)}} \
K_T^{m+1}
\ r^{m+1/2} \ e^{- K_T^2 r^2/2} \ L_n^m(K_T^2 r^2) \,.
\en
Note that the normalizable modes $\Phi_{nJ}(r,T)$ and $\phi_{nJ}(r,T)$
obey the following normalization conditions:
\eq\label{Norm_Cond}
\int\limits_0^\infty dr \, e^{-B_T(r)} \Phi_{mJ}(r,T) \Phi_{nJ}(r,T) =
\int\limits_0^\infty dr \, \phi_{mJ}(r,T) \phi_{nJ}(r,T) = \delta_{mn} \,.
\en
The mode $\Phi_{n}(r,T)$ has the correct behavior in both the
ultraviolet (UV) and infrared (IR) limits:
\eq
\Phi_{nJ}(r,T) \ \sim \ r^{N+L-J}
\ \ {\rm at \ small} \ r\,, \quad\quad
\Phi_{nJ}(r,T) \to 0 \ \ {\rm at \ large} \ r\,.
\en
The above formulas are valid for any set of 
quantum numbers $(n, L, J)$ and number of constituents $N$. 

Finally, in this section we include the effects of finite quark masses 
for low temperatures. Following the ideas developed in our previous 
papers~\cite{Gutsche:2012ez,Gutsche:2017oro} we can readily do it 
for light mesons and tetraquarks. In particular, taking into account 
the $T$-dependence of the light quark condensate parameter $B(T)$, 
we derive the following LO quark mass corrections:  
\eq 
\delta M_\pi^2(T) = 2 \hat{m} B(T) 
\en 
for the pion and 
\eq 
\delta M_K^2(T) = (\hat{m} + m_s) B(T) 
\en 
for the kaon, 
where $\hat{m} = (m_u + m_d)/2$ is the average mass
of $u$ and $d$ quarks, $m_s$ is the strange quark mass. 
The $T$-dependence of $B(T)$ is defined by 
Eqs.~(\ref{defBT}), (\ref{FT1loop}), and (\ref{FT2loop}). 

Light nonstrange $T_{4q}$, single strange $T_{3qs}$, 
double strange $T_{2q2s}$, single nonstrange $T_{3sq}$, 
and strange tetraquark $T_{4s}$ masses get corrections, 
which are simply expressed in terms of the pion and kaon 
masses corrections as~\cite{Gutsche:2017oro}: 
\eq 
& &\delta M_{T_{4q}}^2 \equiv 2 \delta M_\pi^2\,, \quad  
   \delta M_{T_{3qs}}^2 \equiv \delta M_\pi^2 + \delta M_K^2 \nonumber\\
& &\delta M_{T_{2q2s}}^2 \equiv 2 \delta M_K^2 \,, \quad 
   \delta M_{T_{3sq}}^2 \equiv 3 \delta M_K^2 - \delta M_\pi^2 \nonumber\\
& &\delta M_{T_{4s}}^2 \equiv 4 \delta M_K^2 - 2 \delta M_\pi^2 \,.
\en 

\subsection{Hadron form factors at low temperatures}

In this section we derive the results for the form factors of 
hadrons with integer spin (mesons, tetraquarks, 
sixquarks/dibaryons, etc.) at low temperature. 
Following our study in Ref.~\cite{Gutsche:2011vb}, we calculate 
hadron form factors at low temperatures, induced by the coupling 
of AdS fields dual to hadrons with external vector 
AdS fields dual to the electromagnetic field. First, we calculate 
the vector bulk-to-boundary propagator at low temperatures 
using the universal action derived in Eq.~(\ref{action_SM}). 
The corresponding EOM for the Fourier transform of 
the bulk-to-boundary propagator $V(Q,r,T)$, 
in Euclidean metric $Q^2 = -q^2$, reads: 
\eq
\partial_r \biggl( \frac{e^{-\varphi_T(r)}}{r}  \,
\partial_rV(Q,r,T) \biggr) - Q^2 \frac{e^{-\varphi_T(r)}}{r}  \,
V(Q,r,T) = 0 \,.
\en 
This EOM is similar to the EOM for the case of zero
temperature and the only difference is that the temperature
dependence is absorbed in the $T$-dependence of the
dilaton parameter. Therefore, the solution for
the bulk-to-boundary propagator at small temperature
is straightforward~\cite{Grigoryan:2007my}:
\eq\label{VQ_smallT}
V(Q,r,T) &=& \Gamma(1 + a_T) \, U(a_T,0,K_T^2 r^2)
= K_T^2 r^2 \int_0^1 \frac{dx}{(1-x)^2}
\, x^{a_T} \, e^{- K_T^2 r^2 \frac{x}{1-x} } \,, \quad
a_T = \frac{Q^2}{4 K_T^2} \,,
\en
where $\Gamma(n)$ and  $U(x,y,z)$ are the gamma and Tricomi
function, respectively.
Now we can calculate the form factor $F_{nJ}(Q^2,T)$ depending
on the Euclidean momentum squared $Q^2$ for bosonic hadrons
with quantum numbers $(n,J,L)$ and number of constituents $N$
at low temperature. The master formula is
\eq
F_{nJ}(Q^2,T) = \int\limits_0^\infty dr \phi_{nJ}^2(r,T) \,
V(Q,r,T) \,.
\en
Note that at finite temperature the form factor $F_{nJ}(Q^2,T)$ is 
properly normalized with $F_{nJ}(0,T) = 1$ because of $V(0,r,T) = 1$ 
and $\int\limits_0^\infty dr \phi_{nJ}^2(r,T) = 1$. 
Also, $F_{nJ}(Q^2,T)$ has the correct power scaling at large $Q^2$, 
consistent with quark counting rules and independent of the quantum 
numbers $n$ and $J$, while depending on the number of constituents 
$N$ and the orbital momentum $L$: 
\eq 
F_{nJ}(Q^2) \sim \frac{1}{(Q^2)^{m+1}}\,. 
\en 

\clearpage 

\begin{figure} 
\begin{center}
\epsfig{figure=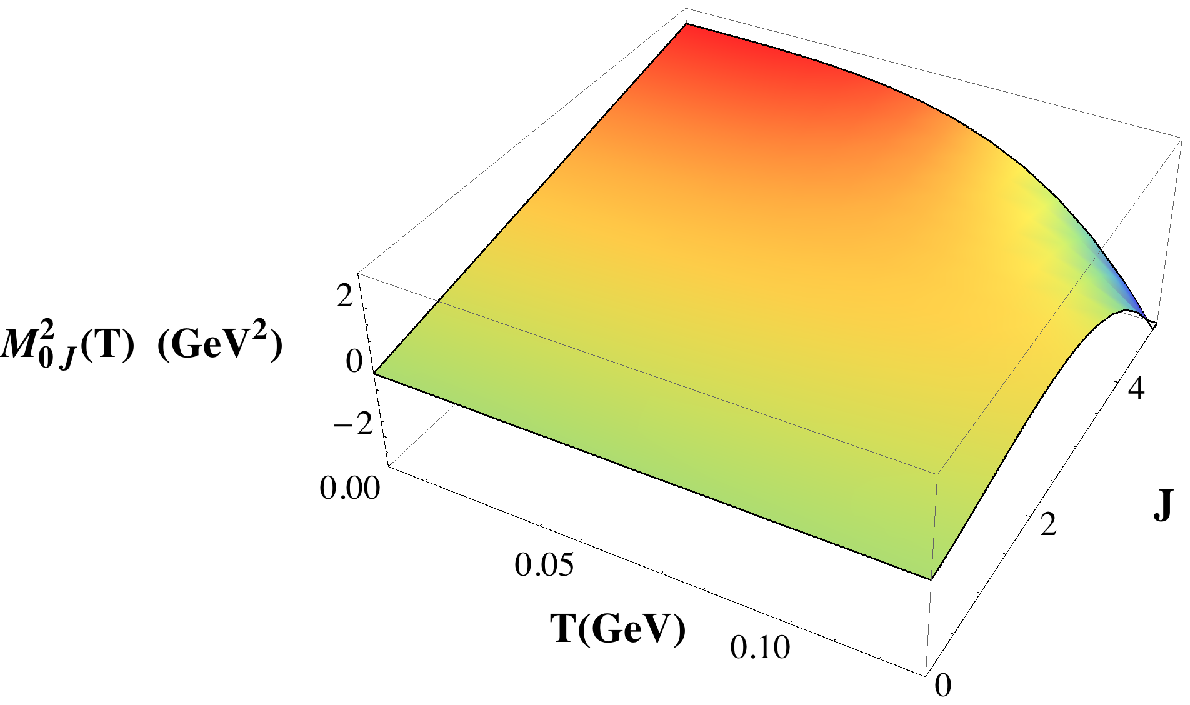,scale=.55}
\quad 
\epsfig{figure=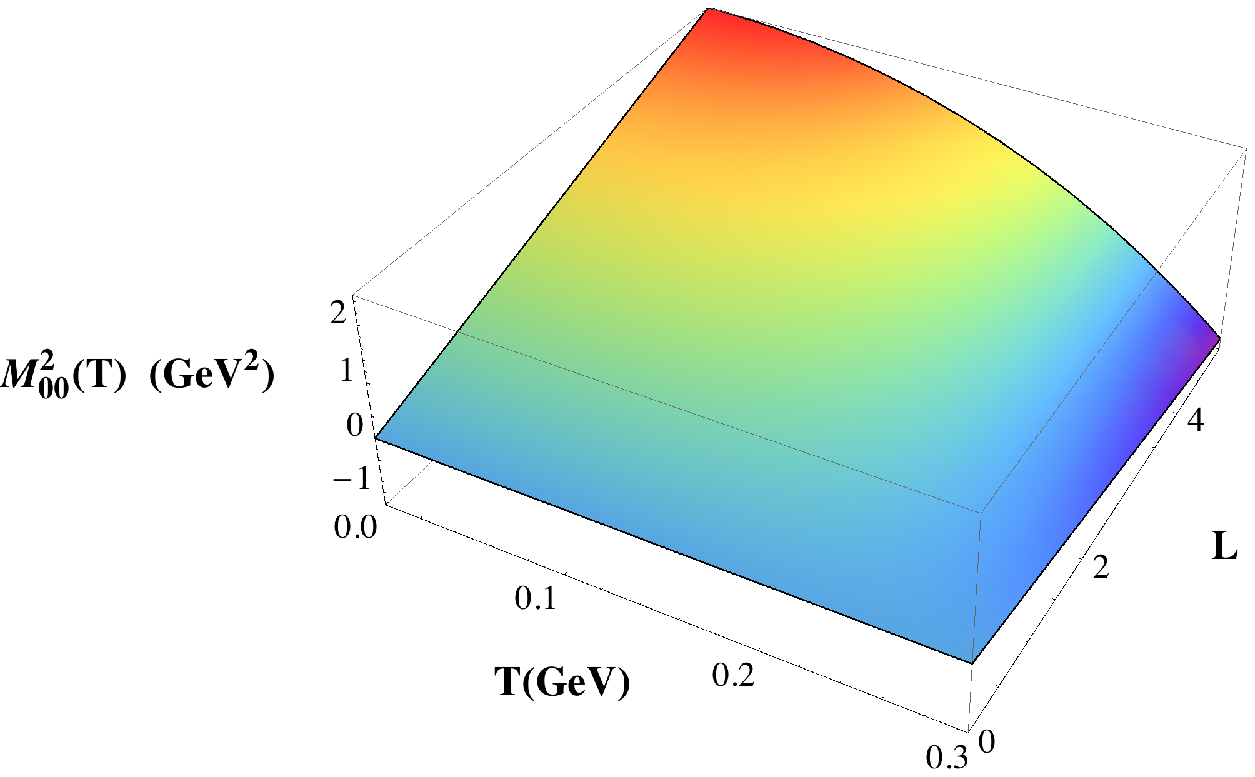,scale=.55}
\end{center}
\noindent
\vspace*{-.75cm}
\centerline{\hspace*{2cm}(a) \hspace*{7cm} (b)}
\caption{$T$-dependence of meson masses: 
(a) $M_{0J}^2(T)$ for $L=0$ and for $J=0,\ldots,5$; 
(b) $M_{00}^2(T)$ for $J=0$ and for $L=0,\ldots,5$.    
\label{fig1}}
\begin{center}
\epsfig{figure=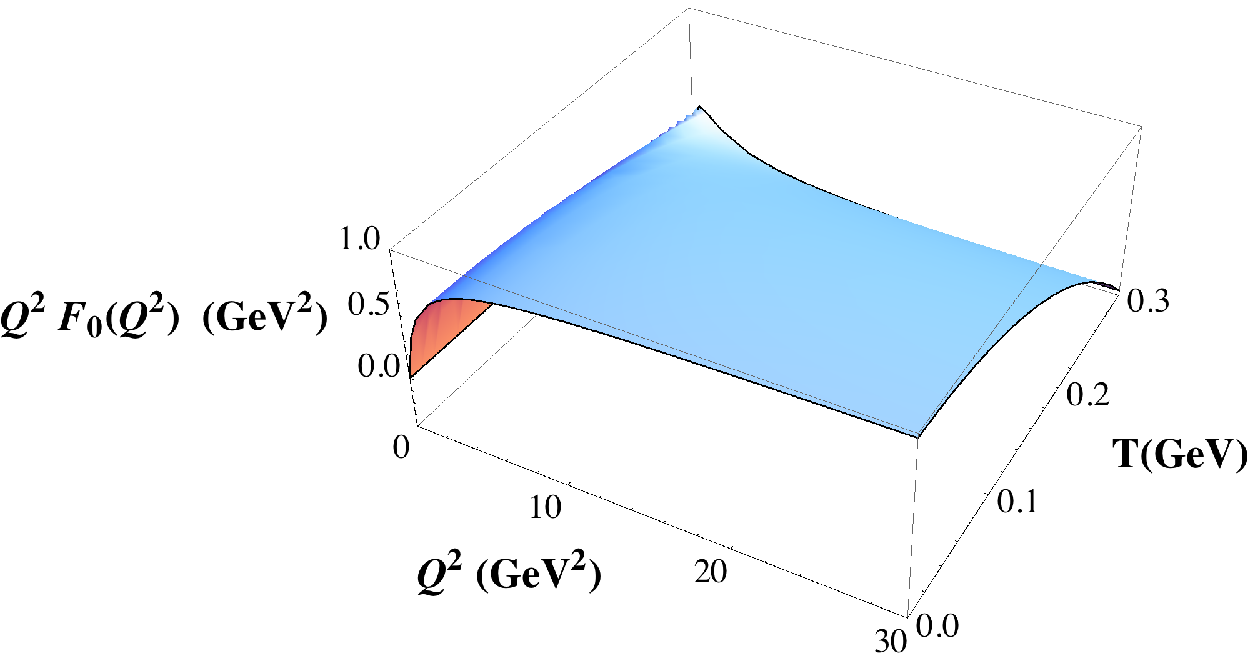,scale=.7}
\end{center}
\noindent
\vspace*{-1cm}
\caption{Temperature dependence of meson form factor 
multiplied by $Q^2$ for $n=0$, $L=0$, and $J=0$. 
\label{fig2}}
\begin{center}
\epsfig{figure=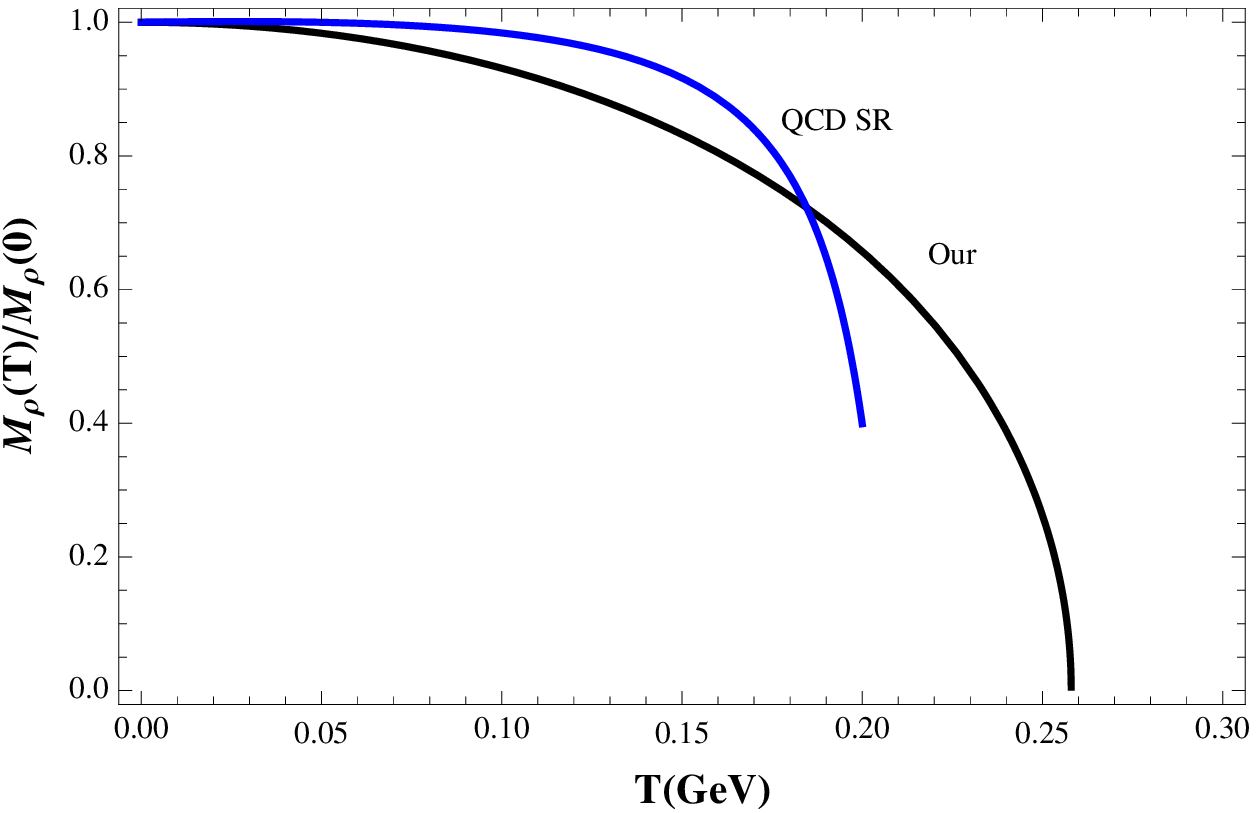,scale=.6}
\end{center}
\noindent
\vspace*{-1cm}
\caption{Comparison of $M_\rho(T)/M_\rho(0)$ 
with result of QCD sum rules~\cite{Ayala:2016vnt} 
\label{fig3}}
\begin{center}
\epsfig{figure=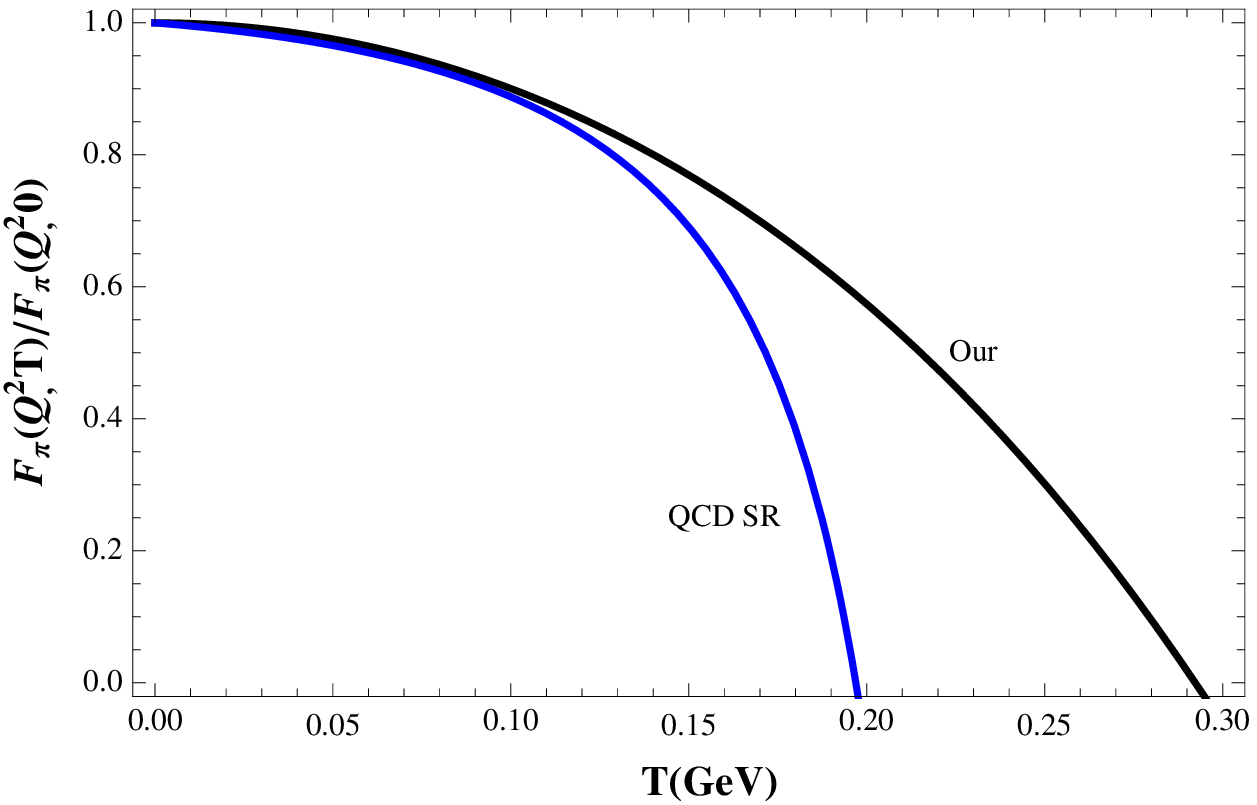,scale=.6}
\end{center}
\noindent
\vspace*{-1cm}
\caption{Comparison of $F_\pi(Q^2,T)/F_\pi(0,T)$ with result 
of QCD sum rules~\cite{Dominguez:1994np} for $Q^2 = 3$ GeV$^2$. 
\label{fig4}}
\end{figure}

\clearpage 

Using Eqs.~(\ref{phi_r0}) and~(\ref{VQ_smallT}) we get for 
the ground state $n=0$ meson 
\eq 
F_{0J}(Q^2,T) = \frac{\Gamma(a_T+1) \, \Gamma(m+2)}{\Gamma(a_T+m+2)}\,.  
\en 
Results for radial excitations with any value for $n$ are readily obtained. 
For example, for the first two radial excitations $n=1$ and $n=2$ 
the form factors are 
\eq
F_{1J}(Q^2,T) &=& \frac{\Gamma(a_T+1) \, \Gamma(m+4)}{\Gamma(a_T+m+4)}
\,+\, a_T (m+1) \frac{\Gamma(a_T+2) \, \Gamma(m+2)}{\Gamma(a_T+m+4)}
\,,\\
F_{2J}(Q^2,T) &=& \frac{\Gamma(a_T+1) \, \Gamma(m+6)}{\Gamma(a_T+m+6)}
\,+\, a_T \frac{\Gamma(a_T+2) \, \Gamma(m+3)}{\Gamma(a_T+m+6)} 
\biggl[ (m+5) (2m+3) + \frac{1}{2} (m+1) a_T (a_T+5) 
\biggr] \,. 
\en 
Next we can perform a small $T$-expansion of the form factors. 
For the ground state form factor we get 
\eq 
F_{0J}(Q^2,T) &=& F_{0J}(Q^2) + \Delta F_{0J}(Q^2)\,, \nonumber\\
F_{0J}(Q^2,0) &=& \frac{\Gamma(a+1) \, \Gamma(m+2)}{\Gamma(a+m+2)}\,, 
\nonumber\\
\Delta F_{0J}(Q^2,T) &=& \rho_T a \frac{\Gamma(a+1) \, 
\Gamma(m+2)}{\Gamma(a+m+2)} \, 
\Big[ \psi(a+m+2) - \psi(a+1) \Big] \,,
\en 
where $a = Q^2/(4 \kappa^2)$ and 
$\psi(n) = \Gamma'(n)/\Gamma(n)$ is the polygamma function. 

\section{Numerical applications}

In this section we present our numerical results for the mass spectrum and  
form factors of mesons at small temperatures and a particular choice for the  
parameter $\alpha = 0.2$ in the dilaton parameter $K^2_T$~(\ref{def_rho}).  
In Fig.~\ref{fig1} we show our results for the temperature dependence of 
meson masses: (a) $M_{0J}^2(T)$ for $n=0$, $L=0$ and for continuous total 
spin $J$ running from 0 to 5 and (b) $M_{00}^2(T)$ for $n=0$,  $J=0$, 
and for continuous orbital angular momentum $L$ running from 0 to 5. 
The temperature dependence of meson form factor multiplied by $Q^2$ 
for $n=0$, $L=0$, and $J=0$ is shown in Fig.~\ref{fig2}. 
A comparison for the $T$ dependence of our results for the $\rho$ meson 
mass and pion form factor $F_\pi(Q^2)$ with results of QCD sum rules 
calculations~\cite{Ayala:2016vnt,Dominguez:1994np} is given in 
Figs.~\ref{fig3} and~\ref{fig4}. We use normalizations 
$M_\rho(T)/M_\rho(0)$ and $F_\pi(Q^2,T)/F_\pi(Q^2,0)$ 
for $Q^2 = 3$ GeV$^2$. 

\section{Summary}

We have proposed a soft-wall AdS/QCD model which implements 
important features of QCD at zero and low temperatures. 
In particular, we showed that the dilaton field, being 
responsible for spontaneous breaking of chiral and conformal 
symmetry, plays an important role in the temperature dependence 
of hadronic properties. The $T$-dependence coincides with the 
one of the quark condensate in QCD. 
In addition to the dilaton we introduce in the action the 
thermal prefactor $e^{-\lambda_T(z)}$, where the thermal function 
$\lambda_T(z)$ contains $z^2$, $z^4$, and $z^6$ terms.  
The $z^4$ term guarantees the gauge invariance and massless ground-state 
pseudoscalar mesons in the chiral limit. The $z^6$ term guarantees the absence 
of power-six terms in the holographic potential. The quadratic 
$z^2$ term gives perturbative contribution to the dilaton field   
and to the restoration of chiral symmetry at a critical temperature $T_c$.
Combining the dilaton 
field with the $z^2$ term in the $\lambda_T(z)$ thermal function, 
we introduced the 
generalized temperature dependent dilaton field. As a consequence 
the thermal behavior of the generalized dilaton is dominated by 
a QCD piece, plus a perturbative term due to the $z^2$ thermal prefactor. 
Using the QCD prediction for the small $T$-dependence of the 
QCD condensate, we predict the thermal behavior of masses 
and form factors of hadrons with integer total angular momentum $J$. 
We present numerical results for the critical 
temperature (when the dilaton field is vanishing), and for 
the $T$-dependence of masses and form factors of mesons.

\begin{acknowledgments}

This work was funded by 
the Carl Zeiss Foundation under Project ``Kepler Center f\"ur Astro- und
Teilchenphysik: Hochsensitive Nachweistechnik zur Erforschung des
unsichtbaren Universums (Gz: 0653-2.8/581/2)'', 
by CONICYT (Chile) under Grants No. 7912010025, 1180232 and PIA/Basal FB0821,
by the Russian Federation program ``Nauka'' (Contract No. 3.6832.2017/8.9), 
by Tomsk State University 
competitiveness improvement program under grant No. 8.1.07.2018, and
by Tomsk Polytechnic University
Competitiveness Enhancement Program (Grant No. VIU-FTI-72/2017).

\end{acknowledgments}

\end{document}